\documentclass[a4paper]{article}

\usepackage{INTERSPEECH2020}

\usepackage{graphicx}
\usepackage{mathtools,amsfonts,amssymb}
\usepackage{algorithm}
\usepackage{algpseudocode}
\usepackage{xcolor}
\usepackage{float}
\usepackage{booktabs, tabularx}
\usepackage{caption}
\usepackage{subcaption}
\usepackage[all]{nowidow}
\usepackage[breaklinks,hidelinks]{hyperref}
\usepackage{soul}

\newcommand{\inpos}{}

\newcommand{\va}{A}
\newcommand{\vh}{H}
\newcommand{\vc}{C}
\newcommand{\vD}{D}
\newcommand{\vw}{w}
\newcommand{\vs}{s}

\newcommand{\ff}{f}

\newcommand{\wave}[1]{\widetilde{\vw}_{#1}}

\newcommand{\dr}{r}
\newcommand{\dd}{d}

\newcommand{\ampltf}[1]{\widetilde{\va}_{#1}\left(t,\ff\right)}
\newcommand{\amplf}[1]{\widetilde{\va}_{#1}\left(\ff\right)}
\newcommand{\respf}[1]{\vh_{#1}\left(\ff\right)}
\newcommand{\magntf}[2][]{\va_{#2}^{#1}\left(t,\ff\right)}

\newcommand{\coeff}[1]{\vc_{#1}\left(\ff\right)}

\newcommand{\sourcecode}{\url{https://github.com/SRPOL-AUI/spectrum-correction}}

\DeclareMathOperator*{\mean}{mean}

\graphicspath{{images/}}

\title{Spectrum Correction: Acoustic Scene Classification \\with Mismatched Recording Devices}
\name{Michał Kośmider}

\address{Samsung R\&D Institute Poland}
\email{m.kosmider@samsung.com}

\begin{document}

\maketitle
\begin{abstract}
  Machine learning algorithms, when trained on audio recordings from a limited set of devices, may not generalize well to samples recorded using other devices with different frequency responses. In this work, a relatively straightforward method is introduced to address this problem. Two variants of the approach are presented. First requires aligned examples from multiple devices, the second approach alleviates this requirement. This method works for both time and frequency domain representations of audio recordings. Further, a relation to standardization and Cepstral Mean Subtraction is analysed. The proposed approach becomes effective even when very few examples are provided. This method was developed during the \textit{Detection and Classification of Acoustic Scenes and Events} (DCASE) 2019 challenge and won the 1st place in the scenario with mismatched recording devices with the accuracy of 75\%. Source code for the experiments can be found online.
\end{abstract}

\section{Introduction}
\label{sec:intro}

Audio machine learning has many sub-fields. Music and speech processing are well established, while other such as event detection, acoustic scene classification (ASC) or behaviour classification are less common. Moreover, the more specific or unusual the task, the harder it is to find publicly available datasets.

When a dataset of audio recordings for a specific purpose is collected, it is usually done with some selected set of recording devices~\cite{ami, dcase}. This is sufficient to test the performance of various algorithms, but might not be enough for practical applications. Models trained using such dataset, when evaluated on samples from the same devices, appear to work well. However, their performance deteriorates when applied on samples from devices that were not in the training set or were underrepresented. This suggests that machine learning algorithms do not generalize across samples recorded using different devices~\cite{sternSignalProcessingRobust1996, transfer-wireless, truc, da-asc,svm-analysis}. In extreme cases, this forces collection of completely new datasets for the targeted devices. Severity of the problem varies between tasks. This effect is clearly visible in acoustic scene classification.

The problem of device mismatch was popularized in the \textit{Detection and Classification of Acoustic Scenes and Events} (DCASE) challenge. In 2018, a task targeting this problem became part of the challenge~\cite{dcase} and has been continued in 2019, which inspired research in this area resulting in various publications~\cite{truc,da-asc,nguyenAcousticSceneClassification2020}. Some of the solutions were based on domain adaptation~\cite{dcase-Eghbal-zadeh2019}, aggressive regularization~\cite{dcase-Eghbal-zadeh2019,dcase-Gao2019} and most were model ensembles~\cite{dcase-Nguyen2018}. 

Independently from DCASE challenge, an approach was proposed to use transfer learning~\cite{transfer-wireless}. The idea was to firstly train the model using a relatively large dataset and then fine tune it using a smaller dataset containing recordings from the targeted devices.

There currently does not appear to be any general approach to address this problem for ASC. Most solutions focus on tailoring specific models to become less sensitive to characteristics of the recording device. On the other hand, in the speech processing community there are multiple well known methods for channel normalization~\cite{junquaRobustnessLanguageSpeech2001, sternSignalProcessingRobust1996, burgetAnalysisFeatureExtraction2007}. However a lot of those methods, such as RASTA~\cite{hermanskyRASTAProcessingSpeech1994}, are leveraging specific characteristics of speech. Others such as Cepstral Mean Subtraction (CMS)~\cite{atalEffectivenessLinearPrediction1974a}, Feature Warping~\cite{pelecanosFeatureWarpingRobust2001} or more recent MVA~\cite{chenMVAProcessingSpeech2007} are relying on quefrency domain, which is not usually applied for ASC and similar tasks. For frequency domain, there exists a recently introduced Per-Channel Energy Normalization (PCEN)~\cite{wangTrainableFrontendRobust2017,lostanlenPerChannelEnergyNormalization2019a}, which additionally reduces noise.
However, noise reduction may not be desirable for ASC, where we would like to retain as much information about the environment as possible. Noise and reverberation are likely to contain information that differentiates various environments.
Because of these considerations, a more universal method could be beneficial.

In this work, a method addressing this problem is described. The approach is straightforward and relies on real-world recordings. It does not influence the design of the architecture of the neural network. This method was designed for ASC, but can be applied to other tasks as well. An initial test of this method was performed during the DCASE 2019 challenge. A submission utilizing this approach won the $1^{st}$ place in the scenario with mismatched recording devices with the accuracy of 75\%~\cite{Komider2019}.

From the acoustic point of view, the presented method appears to bear a similarity to standard microphone calibration techniques~\cite{iec-61094-5}, for example to the secondary calibration of a microphone by the comparison. This method relies on a direct comparison of a microphone with a physically identical microphone having a known calibration and is repeated at each frequency of interest. These methods assume laboratory settings and require expert knowledge, which makes them hard to apply for practitioners without a background in signal processing. Databases of recordings are sometimes provided without explicitly defining the types of devices used in their creation. Also, sometimes the devices are not easily available. The proposed method is simpler, more approachable and could help with the problem of mismatched recording devices.


\section{Proposed Method}
\label{sec:proposed-method}

\subsection{Linear Distortion Model}
\label{ssec:dist-model}

Throughout this paper a linear distortion model will be assumed. Also all amplitudes and responses are positive and real. It can be described in the following way.

Let $\vw_\vs$ be the undistorted waveform representation of some audio signal $\vs$. Let $\wave{\vs,\dd}$ denote the distorted representation of the same signal produced by some device $\dd$. The signal $\vw_\vs$ can be analysed in terms of its amplitude spectrum. Let $\magntf{\vs}\inpos$ denote the amplitude of the frequency $\ff$ at time $t$ in that signal. 
Similarly, we define $\ampltf{\vs,\dd}\inpos$ for signal $\wave{\vs,\dd}$.

The assumption is that the distortion introduced by the device can be approximated using the frequency response of that device $\respf{\dd}\inpos$ as:
\begin{align}
    \label{eq:linear-distortion}
    \ampltf{\vs,\dd} &\approx \respf{\dd} \magntf{\vs} \\
    \label{eq:linear-distortion2}
    &=\respf{\dd} \respf{\vs} \magntf[\prime]{\vs},
\end{align}

\noindent where $\respf{\vs}\inpos$ is the frequency response specific to the signal $\vs$ and independent from the recording device, for example a room impulse response. The amplitude $\magntf[\prime]{\vs}\inpos$ is a remainder after decomposing $\magntf{\vs}$ into $\respf{\vs}$ and everything else. For most tasks, we would want to eliminate both $\respf{\dd}$ and $\respf{\vs}$, but for ASC preserving $\respf{\vs}$ could be desirable, because it is likely to be specific to the environment.

This model is na\"{\i}ve, because it does not take into account for example non-linear nor phase distortions. Nonetheless, it works well in practice.



\subsection{Aligned Spectrum Correction}
\label{ssec:aligned-spectrum-correction}

In this work, the \textit{spectrum correction} (SC) method is introduced. The goal is to transform the distortion produced by some device $\dd$ into the distortion produced by some chosen reference device $\dr$ by matching their spectra. The device $\dd$ can be any known device. When all recording devices are known, this method can be used to transform all samples, so that they appear as if they were recorded using the reference device.

As before, let $\ampltf{\vs,\dd}$ and $\ampltf{\vs,\dr}$ denote amplitudes of frequency $f$ in signals $\wave{\dd}$ and $\wave{\dr}$, recorded using devices $\dd$ and $\dr$, respectively. If device $\dr$ is the reference device, the SC is derived by applying approximation from Equation~\eqref{eq:linear-distortion} twice:
\begin{align}
    \ampltf{\vs,\dr} 
        &\approx \respf{\dr}\magntf{\vs} \\
        &= \frac{\respf{\dd}}{\respf{\dd}}\respf{\dr}\magntf{\vs} \\
        &\approx \frac{\respf{\dr}}{\respf{\dd}}\ampltf{\vs,\dd}
        .
    \label{eq:ar-ax}
\end{align}
Then \textit{correction coefficients} can be defined as:
\begin{equation}
    \coeff{\dd\rightarrow\dr} \coloneqq \frac{\respf{\dr}}{\respf{\dd}}.
    \label{eq:cc}
\end{equation}
Note that the correction coefficients usually cannot be directly computed and have to be estimated from the data, which is explained in Section~\ref{ssec:computing-cc}.
In order to transform a recording from device $\dd$ to look like a recording from device $\dr$, one can simply perform element-wise multiplication
\begin{equation}
    \ampltf{\vs,\dr} \approx \coeff{\dd\rightarrow\dr}\ampltf{\vs,\dd}.
\end{equation}
The correction coefficients can be used to directly convert spectrograms. They can also be used in order to re-synthesise audio using a converted Short Time Fourier Transform (STFT). Alternatively, as shown is Section~\ref{ssec:fir}, using a finite impulse response filter, recordings can be converted directly in the time domain.

\subsection{Computing Correction Coefficients}
\label{ssec:computing-cc}


In order to apply spectrum correction, one needs to compute the correction coefficients. To estimate the coefficients, one can divide
the amplitude spectrum from the reference device by the amplitude spectrum from the source device, element-wise. 
This can be inferred by transforming Equations~\eqref{eq:ar-ax} and \eqref{eq:cc}
\begin{equation}
    \coeff{\dd\rightarrow\dr} 
    = \frac{\respf{\dr}}{\respf{\dd}} 
    \approx \mean_{\vs, t} \frac{\ampltf{\vs,\dr}}{\ampltf{\vs,\dd}}.
\end{equation}
Here $\mean$ refers to the geometric mean, as it appears to be an appropriate choice for averaging scaling factors.
Note that the mean is calculated over both time and examples. 
The significance will be explained in Section~\ref{ssec:env-resp}.

In Equation~\eqref{eq:ar-ax}, it is implicitly assumed that the two devices recorded exactly the same signal. This could be possible only if the devices were in the same point in space at the exactly same moment. In a controlled laboratory environment it is possible to approximately reproduce the same synthetic signal twice. Then by placing the two devices in the exact same location and repeating the experiment, one could obtain such recordings. 

\subsection{Unaligned Spectrum Correction}
\label{ssec:unaligned-spectrum-correction}

Procedure described above requires aligned audio recordings of the same signal. For most applications such examples might not be available. However, this requirement can be easily relaxed. Here the only information that is required is which device produced the recording. 
The trick is to use the geometric mean and its property that
\begin{equation}
    \coeff{\dd\rightarrow\dr}
    \approx \mean_{\vs, t} \frac{\ampltf{\vs,\dr}}{\ampltf{\vs,\dd}}
    =
    \frac{\mean_{\vs, t} \ampltf{\vs,\dr}}{\mean_{\vs, t} \ampltf{\vs,\dd}}
    .
    \label{eq:estimation}
\end{equation}
The averages can be computed independently. Also the recordings do not need to be aligned and can be of different signals. The recorded signals only need to be from similar distributions (requirements are described in the following section).

\subsection{Relation to Standardization}
\label{ssec:simplifications}

It is possible to omit the reference device, when multiplicative constants are of no concern. A recording converted in the following way has a characteristic that does not match any device, but is the same across all of the devices. It does not possess any practical advantages over the unaligned spectrum correction, but it is useful for theoretical analysis and simplified implementation.
Based on the Equation~\eqref{eq:estimation} the following proportion holds
\begin{equation}
    \frac{\mean_{\vs, t} \ampltf{\vs,\dr}}{\mean_{\vs, t} \ampltf{\vs,\dd}}
    \propto \frac{1}{\mean_{\vs, t} \ampltf{\vs,\dd}}
    \approx \coeff{\dd}
    ,
\end{equation}
where $\coeff{\dd}$ are the correction coefficients for device $\dd$.
Then the conversion is performed in the following way
\begin{equation}
    \coeff{\dd}\ampltf{\vs,\dd} \approx \frac{\ampltf{\vs,\dd}}{\mean_{\vs, t} \ampltf{\vs,\dd}}, \quad \forall_{\vs\in \vD_\dd},
\end{equation}
where $\vD_\dd$ is a collection of signals that were recorded using the device $\dd$. Using Equation~\eqref{eq:linear-distortion}, this can be simplified to
\begin{equation}
    \frac{\ampltf{\vs,\dd}}{\mean_{\vs, t} \ampltf{\vs,\dd}}
    = \frac{\respf{\dd}}{\respf{\dd}}\frac{\magntf{\vs}} {\mean_{\vs, t}\magntf{\vs}}.
\end{equation}
The converted recordings are the same for all devices if $\mean_{\vs, t}\magntf{\vs}$ is independent from the device. This is true when devices recorded the same signals. However, it is also true if the signals are not the same, but are from the same distribution.

It turns out that a simple $z$-score standardization can be implemented in such a way as to perform this simplified version. Usual way to perform standardization is to subtract the mean and then divide by the standard deviation. If all averages are assumed to be geometric, then
\begin{align}
    &\log\left(\coeff{\dd}\ampltf{\vs, \dd}\right) \\
    &\approx\log\left(\frac{\ampltf{\vs, \dd}} {\left(\prod_{\vs\in D_\dd}\amplf{\vs, \dd}\right)^{\frac{1}{\left|\vD_\dd\right|}}}\right) \\
    &= \log\ampltf{\vs, \dd} - \frac{1}{\left|\vD_\dd\right|}\sum_{\vs\in D_\dd}\log\amplf{\vs, \dd} \\ 
    &= \log\ampltf{\vs, \dd} - \frac{1}{\left|\vD_\dd\right|T}\sum_{\vs\in D_\dd}\sum_{t}\log\ampltf{\vs, \dd},
    \label{eq:std}
\end{align}
where $\amplf{\vs, \dd}$ denotes $\ampltf{\vs, \dd}$ averaged over time.
This result shows that mean subtraction, when working with log scaled features, such as log-spectrograms or log-melspectrograms is really a special case of the presented method. Note that for this to work the standardization has to be performed for each device and frequency separately. For example average over all examples does not create this effect.
This should not be confused with spectral subtraction~\cite{bollSuppressionAcousticNoise1979}. A somewhat similar result was used for deconvolution~\cite{stockhamBlindDeconvolutionDigital1975}.

\subsection{Relation to Cepstral Mean Subtraction}

An interesting side note is that something resembling Cepstral Mean Subtraction can be derived from Equation~\eqref{eq:std} using linearity of the inverse Fourier transform.
\begin{align}
    &\mathcal{F}^{-1}\left[\log\left(\coeff{\dd}\ampltf{\vs}\right)\right] = \\ 
    &\mathcal{F}^{-1}\left[\log\ampltf{\vs}\right]
    - \frac{1}{\left|\vD_\dd\right|T}\sum_{\vs\in D_\dd}\sum_{t}\mathcal{F}^{-1}\left[\log\ampltf{\vs}\right]. 
    \nonumber
\end{align}
\noindent The usual formulation of CMS computes average only over time for each example separately~\cite{mammoneRobustSpeakerRecognition1996} (see Section~\ref{ssec:env-resp}). 


\subsection{Environment Response}
\label{ssec:env-resp}

A detail worth pointing out is that the mean in all of the four formulations is always computed for a number of recordings, never for the current recording. For example Equation~\eqref{eq:std} has a different effect than
\begin{equation}
    \log\ampltf{\vs,\dd} - \frac{1}{T}\sum_{t}\log\ampltf{\vs,\dd}.
    \label{eq:naive}
\end{equation}
This can be seen after substituting Equation~\eqref{eq:linear-distortion2} into both equations. 
For Equation~\eqref{eq:std} only the $\respf{\dd}$ will cancel out. However, for Equation~\eqref{eq:naive} $\respf{\dd}$ will cancel out, but so will $\respf{\vs}$.
This will result in loss of information about the impulse response of the environment.
Note that the equations have different effect even if $D_\dd$ contains only a single example.

\subsection{Implementation in time domain}
\label{ssec:fir}

The procedure described in Sections~\ref{ssec:aligned-spectrum-correction} and~\ref{ssec:unaligned-spectrum-correction} requires transformation of the signal into the frequency domain. It is completely unnecessary for models taking raw wave-forms as an input. Computing STFT can also be expensive and cumbersome in some situations.

Spectrum correction can be applied directly in the time domain. Correction coefficients can be viewed as gains for frequency bands, then standard techniques from filter design can be used to construct a finite input response (FIR) filter.

\begin{table}[t]
\centering
\caption{Architecture of the network.}
\begin{tabular}{@{}rlll@{}}
\toprule
\textbf{layer} & \textbf{outputs} & \textbf{kernel} & \textbf{stride} \\ \midrule
\textit{Conv2D+ReLU+BN} & 16 & 3 & 1 \\
\textit{Conv2D+ReLU+BN} & 32 & 3 & 2 \\
\textit{Conv2D+ReLU+BN} & 32 & 3 & 1 \\
\textit{Conv2D+ReLU+BN} & 64 & 3 & 2 \\
\textit{Conv2D+ReLU+BN} & 64 & 3 & 1 \\
\textit{AveragePooling} & 64 & - & - \\
\textit{Dense+Softmax} & 10 & - & - \\ \bottomrule
\end{tabular}
\label{tab:model}
\end{table}

\section{Experiments}
\label{sec:experiments}

The experiments for SC were performed based on ASC. The goal of this task is to determine the environment the device is situated in, e.g. that the surroundings of a mobile phone sound like a metro station. The source code for the experiments can be found online (\sourcecode).

\subsection{Dataset}
\label{ssec:dataset}

The \textit{TAU Urban Acoustic Scenes 2019 Mobile}~\cite{dcase} dataset was used for all of the reported experiments.  This was the dataset used in the DCASE 2019 challenge. It consists of $46$ hours of audio recordings divided into equal $10$\,s samples with a sampling rate of $44.1$\,kHz. There are three different recording devices in the dataset. Out of all the recordings, $40$ hours are from a high quality device (device A) and $3$ hours from each of the two mobile devices, Samsung Galaxy S7 (device B) and iPhone SE (device C). The dataset was collected in twelve cities in Europe. 

\subsection{Model and Training}
\label{ssec:model-training}

The classifier was a neural network consisting of five convolutional layers with ReLU activations and batch normalization~\cite{batch-norm} followed by a global average pooling and a single dense layer with a softmax activation, resulting in 71k parameters. The details of this architecture~\cite{Komider2019} can be seen in Table~\ref{tab:model}.

The parameters of the network were optimized using Adadelta~\cite{adadelta} with a cross-entropy loss. The learning rate was reduced by half every time accuracy did not increase for 16 epochs by at least 0.1 of a percentage point. Batch size was set to 64.

Audio samples were transformed to log-melspectrograms in the following way. First, the STFT was computed using a window of 2048 samples and a hop length of 512. To avoid unnecessary computations, spectrum correction was applied at this point (except for the implementation from Equation~\eqref{eq:std}). The resulting output was then converted to mel scale with 256 mel-bins. Next, a logarithm was applied to the amplitudes.
Frequency bins were separately standardized to zero mean and unit variance each, using statistics for the entire training dataset. For the implementation from Equation~\eqref{eq:std} statistics are computed separately for each device.
During training, mixup augmentation~\cite{mixup} was used with alpha parameter set to 0.4. Mixing was performed without the log scaling. 

\begin{figure}[tbh]
    \centering
    \includegraphics[width=\linewidth]{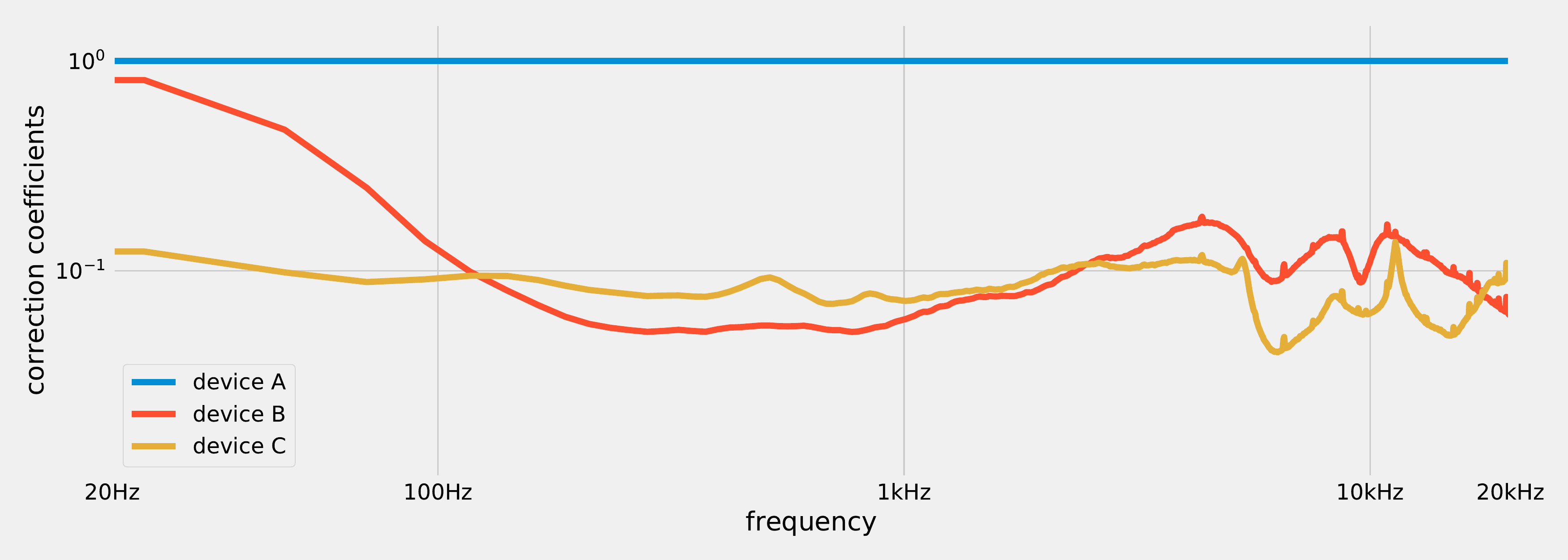}
    \caption{Plot of correction coefficients for devices A, B and C calculated based on 256 examples each. Device A was chosen as the reference device.}
    \label{fig:cc}
\end{figure}

For all experiments device A was chosen as the reference, unless otherwise indicated. Figure~\ref{fig:cc} illustrates the computed coefficients.

\subsection{Time Domain and Reference Device}
\label{ssec:fir-n-ref}

To validate the claim that SC can be applied directly in the time domain, a FIR filter was constructed and tested for various reference devices. The filter was created using the least squares method using 1025 taps. Results for this experiment are shown in Table~\ref{tab:fir-n-ref}. 

\begin{table}[thb]
\centering
\caption{Accuracy computed as macro average for devices A, B and C. Columns represent different choices for the reference device. Spectrum correction is applied both in the frequency domain (STFT) and in the time domain (FIR).}
\begin{tabular}{@{}r|l|lll@{}}
\toprule
\multicolumn{2}{r}{} & \multicolumn{3}{l}{\textbf{reference device}} \\ 
\midrule
\textbf{method}     &\textbf{variant}& \textbf{A} & \textbf{B} & \textbf{C}          \\
\midrule
\textit{aligned}   & STFT & 70.1\%     & 69.4\%     & 69.6\%              \\
\textit{aligned}   & FIR  & 70.6\%     & 70.2\%     & 70.4\%              \\ 
\textit{unaligned} & STFT & 70.2\%     & 69.7\%     & 69.5\%              \\
\textit{unaligned} & FIR  & 70.7\%     & 70.1\%     & 70.5\%              \\ 
\bottomrule
\end{tabular}
\label{tab:fir-n-ref}
\end{table}

The difference in accuracy between models trained with SC applied in the time domain using a FIR filter versus in frequency domain using STFT is negligible. It appears that a FIR filter can be used without a loss in performance.
Also it is clear that the choice of the reference device does not significantly affect the results and that aligned and unaligned variants are indeed equivalent.
On a side note, it is not advisable to use the simplified formulation from Section~\ref{ssec:simplifications} to create the filters (due to the range of the gains).

\subsection{Required Size of the Dataset}
\label{ssec:num-examples}

Various numbers of recordings can be used to compute the correction coefficients. It is interesting to see how many are actually needed. Performance of models trained using the correction coefficients computed using a limited number of examples is shown in Table~\ref{tab:accuracy-num-examples}. It is surprising to see that the accuracy does not change with the increasing number of recordings as much as one would expect.

\begin{table}[htb]
\centering
\caption{Accuracy for devices A, B and C with respect to the number of recordings used to compute the correction coefficients. Models were trained for 100 epochs.}
\begin{tabular}{@{}r|llll@{}}
\toprule
                & \multicolumn{4}{l}{\textbf{number of recordings}}         \\ \midrule
\textbf{device} & \textbf{1} & \textbf{4} & \textbf{16} & \textbf{128} \\ \midrule
\textit{A}      & 72.2\%     & 73.1\%     & 72.4\%      & 72.9\%       \\
\textit{B}      & 65.6\%     & 63.7\%     & 64.1\%      & 64.6\%       \\
\textit{C}      & 69.0\%     & 70.0\%     & 69.6\%      & 71.1\%       \\ \bottomrule
\end{tabular}
\label{tab:accuracy-num-examples}
\end{table}

\subsection{Comparison with Other Methods}
\label{ssec:with-n-without}

An important question is how models trained with and without SC compare to models trained using other methods, in terms of their accuracy. Results for those experiments are shown in Table~\ref{tab:with-n-without}.
For all methods, best found configurations were reported. 
RASTA used alternative with pole at $0.98$.
For PCEN~(a) $T=0.01$\,s, $\varepsilon=10^{-6}$, $\alpha=0.95$, $\delta=2$ and $r = 0.5$.
For PCEN~(b) $T=0.2$\,s, $\alpha=1.0$. The two variants for PCEN trade off accuracy between device A and the mobile devices.
Both implementations use Librosa~\cite{mcfeeLibrosaAudioMusic2015}. RASTA implementation is based on MATLAB implementation~\cite{ellisPLPRASTAMFCC2005}. CMS is performed in frequency domain (see Equation~\eqref{eq:naive}). 

\begin{table}[thb]
\centering
\caption{Accuracy for devices A, B and C for models trained with and without SC and using other methods. Models were trained for 100 epochs.}
\begin{tabular}{@{}rllll@{}}
\toprule
                             & \multicolumn{4}{l}{\textbf{source device}}               \\ \midrule
\textbf{method}              & \textbf{A}          & \textbf{B}      & \textbf{C} & \textbf{mean}       \\ \midrule
\textit{RASTA}               & 60.2\% & 59.7\% & 60.9\% & 60.1\%    \\
\textit{PCEN (a)}            & 69.5\% & 55.4\% & 57.0\% & 60.6\%    \\
\textit{PCEN (b)}            & 64.0\% & 58.6\% & 64.4\% & 62.3\%    \\
\textit{--}                  & 71.6\% & 59.2\% & 61.2\% & 64.0\%    \\
\textit{CMS}                 & 63.9\% & 61.4\% & 67.1\% & 64.2\%    \\
\textit{SC aligned}          & \textbf{73.4\%} & \textbf{65.4\%} & \textbf{71.8\%} & \textbf{70.2\%}    \\
\textit{SC unaligned}        & \textbf{73.1\%} & \textbf{65.9\%} & \textbf{71.4\%} & \textbf{70.4\%}    \\
\textit{SC mean subtraction} & \textbf{73.2\%} & \textbf{65.8\%} & \textbf{72.2\%} & \textbf{70.3\%}    \\
\bottomrule
\end{tabular}
\label{tab:with-n-without}
\end{table}

Methods from speech processing do not appear to perform well on ASC.
However, CMS provided improvement for the mobile devices, nonetheless it deteriorated the accuracy on the main high-quality device A. This was consistent with the expectations~\cite{mammoneRobustSpeakerRecognition1996}. In contrast all implementations of SC resulted in significantly better accuracy for mobile devices. Interestingly, accuracy for device A also increased. It is likely caused by more efficient use of recordings from mobile devices.

\section{Summary}
\label{sec:summary}

In this paper, a new approach to address learning with mismatched recording devices was proposed. It was experimentally shown that using just a few examples, it is possible to significantly improve the accuracy of a model. Particularly, accuracy for two mobile devices increased from 59\% to 66\% and from 61\% to 72\%, when SC was applied. It also outperforms classical methods such as RASTA or CMS, when applied on ASC. 
This method can be applied both in the frequency domain using the STFT, and in the time domain using FIR filters. 

\section{Acknowledgements}

I thank Zuzanna Kwiatkowska, Sławomir Kapka, Kornel Jankowski, Krzysztof Rykaczewski, Mateusz Matuszewski, Michał Łopuszyński and Piotr Masztalski for review of this paper and Jakub Tkaczuk for help with administrative duties.


\bibliographystyle{IEEEtran}
\bibliography{mybib, bib-sc}

\begin{thebibliography}{10}
\providecommand{\url}[1]{#1}
\csname url@samestyle\endcsname
\providecommand{\newblock}{\relax}
\providecommand{\bibinfo}[2]{#2}
\providecommand{\BIBentrySTDinterwordspacing}{\spaceskip=0pt\relax}
\providecommand{\BIBentryALTinterwordstretchfactor}{4}
\providecommand{\BIBentryALTinterwordspacing}{\spaceskip=\fontdimen2\font plus
\BIBentryALTinterwordstretchfactor\fontdimen3\font minus
  \fontdimen4\font\relax}
\providecommand{\BIBforeignlanguage}[2]{{%
\expandafter\ifx\csname l@#1\endcsname\relax
\typeout{** WARNING: IEEEtran.bst: No hyphenation pattern has been}%
\typeout{** loaded for the language `#1'. Using the pattern for}%
\typeout{** the default language instead.}%
\else
\language=\csname l@#1\endcsname
\fi
#2}}
\providecommand{\BIBdecl}{\relax}
\BIBdecl

\bibitem{ami}
T.~{Hain}, L.~{Burget}, J.~{Dines}, P.~N. {Garner}, F.~{Grezl}, A.~E.
  {Hannani}, M.~{Huijbregts}, M.~{Karafiat}, M.~{Lincoln}, and V.~{Wan},
  ``Transcribing meetings with the amida systems,'' \emph{IEEE Transactions on
  Audio, Speech, and Language Processing}, vol.~20, no.~2, pp. 486--498, Feb
  2012.

\bibitem{dcase}
\BIBentryALTinterwordspacing
A.~Mesaros, T.~Heittola, and T.~Virtanen, ``A multi-device dataset for urban
  acoustic scene classification,'' in \emph{Proceedings of the Detection and
  Classification of Acoustic Scenes and Events 2018 Workshop (DCASE2018)},
  November 2018, pp. 9--13. [Online]. Available:
  \url{https://arxiv.org/abs/1807.09840}
\BIBentrySTDinterwordspacing

\bibitem{sternSignalProcessingRobust1996}
R.~M. Stern, A.~Acero, F.-H. Liu, and Y.~Ohshima,
  ``\BIBforeignlanguage{en}{Signal {Processing} for {Robust} {Speech}
  {Recognition}},'' in \emph{\BIBforeignlanguage{en}{Automatic {Speech} and
  {Speaker} {Recognition}: {Advanced} {Topics}}}, ser. The {Kluwer}
  {International} {Series} in {Engineering} and {Computer} {Science}, C.-H.
  Lee, F.~K. Soong, and K.~K. Paliwal, Eds.\hskip 1em plus 0.5em minus
  0.4em\relax Boston, MA: Springer US, 1996, pp. 357--384.

\bibitem{transfer-wireless}
G.~Gosztolya and T.~Grósz, ``Domain adaptation of deep neural networks for
  automatic speech recognition via wireless sensors,'' \emph{Journal of
  Electrical Engineering}, vol.~67, 04 2016.

\bibitem{truc}
T.~{Nguyen} and F.~{Pernkopf}, ``Acoustic scene classification with mismatched
  recording devices using mixture of experts layer,'' in \emph{2019 IEEE
  International Conference on Multimedia and Expo (ICME)}, July 2019, pp.
  1666--1671.

\bibitem{da-asc}
\BIBentryALTinterwordspacing
S.~Gharib, K.~Drossos, E.~Cakir, D.~Serdyuk, and T.~Virtanen, ``{U}nsupervised
  {A}dversarial {D}omain {A}daptation for {A}coustic {S}cene
  {C}lassification,'' in \emph{Detection and Classification of Acoustic Scenes
  and Events}.\hskip 1em plus 0.5em minus 0.4em\relax Tampere University of
  Technology, 2018. [Online]. Available: \url{https://arxiv.org/abs/1808.05777}
\BIBentrySTDinterwordspacing

\bibitem{svm-analysis}
B.~{Ma}, H.~M. {Meng}, and M.~{Mak}, ``Effects of device mismatch, language
  mismatch and environmental mismatch on speaker verification,'' in \emph{2007
  IEEE International Conference on Acoustics, Speech and Signal Processing -
  ICASSP '07}, vol.~4, April 2007, pp. IV--301--IV--304.

\bibitem{nguyenAcousticSceneClassification2020}
T.~Nguyen, F.~Pernkopf, and M.~Kosmider, ``Acoustic {Scene} {Classification}
  for {Mismatched} {Recording} {Devices} {Using} {Heated}-{Up} {Softmax} and
  {Spectrum} {Correction},'' in \emph{{ICASSP} 2020 - 2020 {IEEE}
  {International} {Conference} on {Acoustics}, {Speech} and {Signal}
  {Processing} ({ICASSP})}, May 2020, pp. 126--130, iSSN: 2379-190X.

\bibitem{dcase-Eghbal-zadeh2019}
H.~Eghbal-zadeh, K.~Koutini, and G.~Widmer, ``Acoustic scene classification and
  audio tagging with receptive-field-regularized {CNNs},'' DCASE2019 Challenge,
  Tech. Rep., June 2019.

\bibitem{dcase-Gao2019}
W.~Gao and M.~McDonnell, ``Acoustic scene classification using deep residual
  networks with late fusion of separated high and low frequency paths,''
  DCASE2019 Challenge, Tech. Rep., June 2019.

\bibitem{dcase-Nguyen2018}
T.~Nguyen and F.~Pernkopf, ``Acoustic scene classification using a
  convolutional neural network ensemble and nearest neighbor filters,''
  DCASE2018 Challenge, Tech. Rep., September 2018.

\bibitem{junquaRobustnessLanguageSpeech2001}
\BIBentryALTinterwordspacing
J.-C. Junqua, G.~van Noord, N.~Ide, and J.~Véronis, Eds.,
  \emph{\BIBforeignlanguage{en}{Robustness in {Language} and {Speech}
  {Technology}}}, ser. Text, {Speech} and {Language} {Technology}.\hskip 1em
  plus 0.5em minus 0.4em\relax Dordrecht: Springer Netherlands, 2001, vol.~17.
  [Online]. Available: \url{http://link.springer.com/10.1007/978-94-015-9719-7}
\BIBentrySTDinterwordspacing

\bibitem{burgetAnalysisFeatureExtraction2007}
L.~Burget, P.~Matejka, P.~Schwarz, O.~Glembek, and J.~H. Cernocky, ``Analysis
  of {Feature} {Extraction} and {Channel} {Compensation} in a {GMM} {Speaker}
  {Recognition} {System},'' \emph{IEEE Transactions on Audio, Speech, and
  Language Processing}, vol.~15, no.~7, pp. 1979--1986, Sep. 2007.

\bibitem{hermanskyRASTAProcessingSpeech1994}
H.~Hermansky and N.~Morgan, ``{RASTA} processing of speech,'' \emph{IEEE
  Transactions on Speech and Audio Processing}, vol.~2, no.~4, pp. 578--589,
  Oct. 1994.

\bibitem{atalEffectivenessLinearPrediction1974a}
\BIBentryALTinterwordspacing
B.~S. Atal, ``\BIBforeignlanguage{en}{Effectiveness of linear prediction
  characteristics of the speech wave for automatic speaker identification and
  verification},'' \emph{\BIBforeignlanguage{en}{The Journal of the Acoustical
  Society of America}}, vol.~55, no.~6, pp. 1304--1312, Jun. 1974, publisher:
  Acoustical Society of America. [Online]. Available:
  \url{http://asa.scitation.org/doi/10.1121/1.1914702}
\BIBentrySTDinterwordspacing

\bibitem{pelecanosFeatureWarpingRobust2001}
\BIBentryALTinterwordspacing
J.~Pelecanos and S.~Sridharan, ``\BIBforeignlanguage{en}{Feature {Warping} for
  {Robust} {Speaker} {Verification}},'' in
  \emph{\BIBforeignlanguage{en}{Proceedings of 2001 {A} {Speaker} {Odyssey}:
  {The} {Speaker} {Recognition} {Workshop}}}.\hskip 1em plus 0.5em minus
  0.4em\relax Crete, Greece: European Speech Communication Association, 2001,
  pp. 213--218, conference Name: 2001 A Speaker Odyssey: The Speaker
  Recognition Workshop Meeting Name: 2001 A Speaker Odyssey: The Speaker
  Recognition Workshop. [Online]. Available:
  \url{https://eprints.qut.edu.au/10408/}
\BIBentrySTDinterwordspacing

\bibitem{chenMVAProcessingSpeech2007}
C.-P. Chen and J.~A. Bilmes, ``{MVA} {Processing} of {Speech} {Features},''
  \emph{IEEE Transactions on Audio, Speech, and Language Processing}, vol.~15,
  no.~1, pp. 257--270, Jan. 2007, conference Name: IEEE Transactions on Audio,
  Speech, and Language Processing.

\bibitem{wangTrainableFrontendRobust2017}
Y.~Wang, P.~Getreuer, T.~Hughes, R.~F. Lyon, and R.~A. Saurous, ``Trainable
  {Frontend} {For} {Robust} and {Far}-{Field} {Keyword} {Spotting},'' in
  \emph{Proc. {IEEE} {ICASSP} 2017}, New Orleans, LA, 2017.

\bibitem{lostanlenPerChannelEnergyNormalization2019a}
\BIBentryALTinterwordspacing
V.~Lostanlen, J.~Salamon, M.~Cartwright, B.~McFee, A.~Farnsworth, S.~Kelling,
  and J.~P. Bello, ``\BIBforeignlanguage{en}{Per-{Channel} {Energy}
  {Normalization}: {Why} and {How}},'' \emph{\BIBforeignlanguage{en}{IEEE
  Signal Processing Letters}}, vol.~26, no.~1, pp. 39--43, Jan. 2019. [Online].
  Available: \url{https://ieeexplore.ieee.org/document/8514023/}
\BIBentrySTDinterwordspacing

\bibitem{Komider2019}
\BIBentryALTinterwordspacing
M.~Kośmider, ``Calibrating neural networks for secondary recording devices,''
  DCASE2019 Challenge, Tech. Rep., Jun. 2019. [Online]. Available:
  \url{http://dcase.community/documents/challenge2019/technical_reports/DCASE2019_Kosmider_61.pdf}
\BIBentrySTDinterwordspacing

\bibitem{iec-61094-5}
\BIBentryALTinterwordspacing
``\BIBforeignlanguage{en}{Electroacoustics - measurement microphones - part 5:
  Methods for pressure calibration of working standard microphones by
  comparison},'' International Electrotechnical Commission, Geneva, CH,
  International Standard IEC 61094-5:2016, 2016. [Online]. Available:
  \url{https://webstore.iec.ch/publication/24988}
\BIBentrySTDinterwordspacing

\bibitem{bollSuppressionAcousticNoise1979}
S.~Boll, ``Suppression of acoustic noise in speech using spectral
  subtraction,'' \emph{IEEE Transactions on Acoustics, Speech, and Signal
  Processing}, vol.~27, no.~2, pp. 113--120, Apr. 1979, conference Name: IEEE
  Transactions on Acoustics, Speech, and Signal Processing.

\bibitem{stockhamBlindDeconvolutionDigital1975}
T.~Stockham, T.~Cannon, and R.~Ingebretsen, ``Blind deconvolution through
  digital signal processing,'' \emph{Proceedings of the IEEE}, vol.~63, no.~4,
  pp. 678--692, Apr. 1975.

\bibitem{mammoneRobustSpeakerRecognition1996}
R.~Mammone, X.~Zhang, and R.~Ramachandran, ``Robust speaker recognition: {A}
  feature-based approach,'' \emph{Signal Processing Magazine, IEEE}, vol.~13,
  p.~58, Oct. 1996.

\bibitem{batch-norm}
\BIBentryALTinterwordspacing
S.~Ioffe and C.~Szegedy, ``Batch normalization: Accelerating deep network
  training by reducing internal covariate shift,'' in \emph{Proceedings of the
  32nd International Conference on Machine Learning}, ser. Proceedings of
  Machine Learning Research, F.~Bach and D.~Blei, Eds., vol.~37.\hskip 1em plus
  0.5em minus 0.4em\relax Lille, France: PMLR, 07--09 Jul 2015, pp. 448--456.
  [Online]. Available: \url{http://proceedings.mlr.press/v37/ioffe15.html}
\BIBentrySTDinterwordspacing

\bibitem{adadelta}
M.~D. Zeiler, ``Adadelta: an adaptive learning rate method,'' \emph{arXiv
  preprint arXiv:1212.5701}, 2012.

\bibitem{mixup}
\BIBentryALTinterwordspacing
H.~Zhang, M.~Cisse, Y.~N. Dauphin, and D.~Lopez-Paz, ``mixup: Beyond empirical
  risk minimization,'' in \emph{International Conference on Learning
  Representations}, 2018. [Online]. Available:
  \url{https://openreview.net/forum?id=r1Ddp1-Rb}
\BIBentrySTDinterwordspacing

\bibitem{mcfeeLibrosaAudioMusic2015}
B.~McFee, C.~Raffel, D.~Liang, D.~Ellis, M.~Mcvicar, E.~Battenberg, and
  O.~Nieto, ``librosa: {Audio} and music signal analysis in python,'' in
  \emph{Proceedings of the 14th python in science conference}, Jan. 2015, pp.
  18--24.

\bibitem{ellisPLPRASTAMFCC2005}
\BIBentryALTinterwordspacing
D.~P.~W. Ellis, ``{PLP} and {RASTA} (and {MFCC}, and inversion) in {Matlab},''
  2005. [Online]. Available:
  \url{http://www.ee.columbia.edu/~dpwe/resources/matlab/rastamat/}
\BIBentrySTDinterwordspacing

\end{thebibliography}

\end{document}